\documentstyle[namedreferences]{kluwer}
\runningtitle{NUCLEAR ASPECTS OF THE SOLAR NEUTRINO PROBLEM}
\runningauthor{H. OBERHUMMER ET AL.}

\begin{opening}

\title{NUCLEAR ASPECTS OF THE SOLAR NEUTRINO PROBLEM}

\author{H. \surname{OBERHUMMER}}
\author{A.N. \surname{IVANOV}\thanks{%
Permanent Address:
State Technical University, Department of Theoretical
Physics, 195251 St. Petersburg, Russian Federation.}}
\author{N.I. \surname{TROITSKAYA$^*$}}
\author{M. \surname{FABER}}
\institute{Institut f\"ur Kernphysik, Technische Universit\"at Wien,\\
Wiedner Hauptstr. 8-10, A-1040 Vienna, Austria}
\end{opening}

\begin{document}

\begin{abstract}
The present status of the nuclear reaction rates determining
the solar neutrino flux is discussed. This includes the reaction rates
for the two branching ratios of the three pp-chains
involving the reactions $^3$He($^3$He,2p)$^4$He
and $^3$He($^4$He,$\gamma$)$^7$Be for the first branching,
and $^7$Be(e$^-$,$\nu_{\rm e}$)$^7$Li and $^7$Be(p,$\gamma$)$^8$B
for the second branching.
Mainly we will concentrate on the basic nuclear reaction
p + p $\to$ D + $e^{\,+}$ + $\nu_{\rm e}$ of
the pp-chains.
We use a relativistic field theory model of the deuteron
to calculate the low-energy cross section
for this reaction. The theoretical prediction of the
cross section obtained is about 2.9 times larger than given in the conventional
potential approach.
The consequences of this new reaction rate for the solar neutrino
problem will be presented and discussed.
\end{abstract}
  

\section{Introduction}

The solar neutrino problem has its origin in the discrepancy between the 
observed terrestrial neutrino fluxes
of the neutrinos produced by nuclear reactions in the solar core 
and the predicted neutrino
fluxes by standard solar models (SSM's)~\cite{Bahcall89,Bahcall95,Castellani97}.
The observation of the
solar neutrinos is of decisive importance, because it allows us contrary
to the observed solar electromagnetic radiation
to look into the nuclear active zone in the core
of the sun, since the solar mantle is practically transparent
to neutrinos.

The observed neutrino fluxes are substantially less than predicted
by SSM's. Roughly speaking the Homesteak
chlorine-detector~\cite{Davis68,Cleveland95},
the neutrino-electron scattering detector
KAMIOKANDE~\cite{Hirata89,KAMIOKANDE95},
and the gallium detectors GALLEX~\cite{GALLEX94,GALLEX96}
and SAGE~\cite{Abazov91,SAGE96}
observe approximately the fractions 3/10, 4/10, and 5/10 of the predicted
solar neutrino flux, respectively (see Table I). Recently,
the reliability of the measuring method of the GALLEX collaboration was
tested with an artificial neutrino source, resembling
the solar neutrino spectrum that
was inserted in the gallium tank~\cite{Hampel96}.
The measured neutrino flux was $(92 \pm 8)$\,\% of the 
expected value. This result improves significantly the
credibility of the obtained values for the measured solar neutrino flux
in the GALLEX detector.

The present values and their 1$\sigma$ errors
of the different neutrino detectors are shown in Table I.
There are three different nuclear reactions in the pp-chains
of the sun emitting the bulk of solar neutrinos (Fig.~1):\\
(i) Two-proton fusion: p + p $\to$ D + $e^{\,+}$ + $\nu_{\rm e}$
(pp-neutrinos)\\
(ii) Electron capture by $^7$Be: $^7$Be(e$^-$,$\nu_{\rm e}$)$^7$Li
(Be-neutrinos)\\
(iii) Beta-decay by $^8$B: $^8$B(,e$^+$$\nu_{\rm e}$)$^8$Be
(B-neutrinos).\\
The flux of other solar neutrino sources (pep- and CNO-neutrinos) is much
weaker.

\begin{table}
\caption{The present different neutrino detectors with their detection
reactions, threshold energies and the detected neutrino
flux. In the last column the the result obtained using
a standard solar model (SSM) are given.}
\begin{center}
\begin{tabular}{ccccc}\hline
Neutrino detector & threshold & Observed & SSM$^{\rm c}$ \\ 
detection reaction & energy$^{\rm a}$ & result$^{\rm b}$ & result$^{\rm
b}$\\\hline                                       
Homesteak & 0.814 & $2.55 \pm 0.17 \pm 0.18$$^{\rm d}$ & $9.3^{+1.2}_{-1.4}$\\
$\nu$ + $^{37}$Cl $\rightarrow$ e$^-$ + $^{37}$Ar &&&\\
KAMIOKANDE & 7 & $2.73 \pm 0.17 \pm 0.34$$^{\rm e}$ & $6.62
\times \left(1.00^{+0.14}_{-0.17}\right)$\\
$\nu$ + e$^-$ $\rightarrow$ $\nu$ + e$^-$ &&&\\
GALLEX & 0.233 & $69.7^{+7.8}_{-8.1}$$^{\rm f}$ & $137^{+8}_{-7}$\\
$\nu$ + $^{71}$Ga $\rightarrow$ e$^-$ + $^{71}$Ge &&&\\
SAGE & 0.233 & $69\pm10^{+5}_{-7}$$^{\rm g}$ & $137^{+8}_{-7}$\\
$\nu$ + $^{71}$Ga $\rightarrow$ e$^-$ + $^{71}$Ge &&&\\\hline
\end{tabular}
\end{center}

\noindent
{\footnotesize $^{\rm a}$ Energy in MeV\\
$^{\rm b}$ in 10$^6$\,cm$^{-2}$\,s$^{-1}$ for KAMIOKANDE;
in SNU (10$^{-36}$ captures per target atom and seconds) for the others\\
$^{\rm c}$\citeauthor{Bahcall95}, \citeyear{Bahcall95}\\ 
$^{\rm d}$ \citeauthor{Cleveland95}, \citeyear{Cleveland95}\\
$^{\rm e}$ \citeauthor{KAMIOKANDE95}, \citeyear{KAMIOKANDE95}\\
$^{\rm f}$ \citeauthor{GALLEX96}, \citeyear{GALLEX96}\\
$^{\rm g}$ \citeauthor{SAGE96}, \citeyear{SAGE96}
}
\end{table}

The spectrum of the solar neutrinos as predicted
by SSM's is depicted in Fig.~2. In this figure 
also the thresholds of the different
neutrino detectors are indicated. The pp-neutrinos can
only be detected by the gallium detectors, whereas
the neutrino-electron scattering detector can only
detect the B-neutrinos.

In this article we investigate in Sect.~2 the
nuclear reactions determining mainly the solar neutrino flux.
In Sect.~3 we discuss the new reaction rate
for the reaction p + p $\to$ D + $e^{\,+}$ + $\nu_{\rm e}$
using a relativistic field theory model of the deuteron.
Some consequences of this new reaction rate for the solar neutrino flux
are presented and discussed in Sect.~4.

\section{Nuclear reactions determining the solar neutrino flux}

The nuclear reactions relevant for the solar neutrino flux
are:\\
(i) The basic nuclear reaction in the solar core:
p + p $\to$ D + $e^{\,+}$ + $\nu_{\rm e}$\\
(ii) The reactions determining the branching ratio between
the ppI- and (ppII+ppIII)-chains (see Fig.~1):
$^3$He($^3$He,2p)$^4$He and $^3$He($^4$He,$\gamma$)$^7$Be\\
(iii) The reactions determining the branching ratio between
the ppII- and ppIII-chains (see Fig.~1):
$^7$Be(e$^-$,$\nu_{\rm e}$)$^7$Li, and $^7$Be(p,$\gamma$)$^8$B.

The low-energy electroweak model and non-relativistic approaches, having
been applied 
to the computation of the contribution of strong interactions to the matrix
element of 
the p + p $\to$ D + W$^+$ transition~\cite{Bethe39,Bahcall92,Kamionkowski94}
do not leave room for a substantial change of 
the cross section magnitude. As has been noted by
\citeauthor{Castellani97}, \citeyear{Castellani97} they differ from the 
mean value by no more than 3\,\%.

The values of the astrophysical S-factors for the reactions
$^3$He($^3$He,2p)$^4$He
and $^3$He($^4$He,$\gamma$)$^7$Be
are estimated to be known within about $\pm$4\,\%~\cite{Castellani97}.
Recently, it was possible to measure the reaction $^3$He($^3$He,2p)$^4$He
down to the energies of the solar Gamow peak~\cite{Arpesella96}. This
reduced the uncertainty of the
astrophysical S-factor for this reaction by about another factor of 2.
Furthermore,
no sign of a hypothetical resonance at the energies of the solar Gamow peak
was found for this reaction.

The $^7$Be electron capture rate $^7$Be(e$^-$,$\nu_{\rm e}$)$^7$Li
has recently been investigated by
\citeauthor{Gruzinov97}, \citeyear{Gruzinov97}. The total theoretical
uncertainty in the electron capture rate under solar conditions has
been found to be about 2\,\%. The largest uncertainty is the
value of the astrophysical S-factor of $^7$Be(p,$\gamma$)$^8$B.
It is estimated to be known only within approximately
$\pm 10$\,\% \cite{Castellani97}. However, because the
ppIII-chain is so weak compared to the ppII-chain (see Fig.~1), only
the $^8$B solar neutrino flux is significantly changed by this uncertainty.

Summarizing, by using the above discussed uncertainties
in the astrophysical S-factors and
reaction rates there seems to be no
possibility to relax the solar neutrino problem
in the framework of pure nuclear physics. This has already been noticed
before by many authors.

\section{A new reaction rate for p + p $\to$ D + $e^{\,+}$ + $\nu_{\rm e}$}

A substantial enhancement of the reaction rate for
the two-proton fusion p + p $\to$ D + e$^+$ + $\nu_{\rm e}$ 
by a factor of about 2.9 with respect to the potential approach has been found
recently~\cite{Ivanov97}
within a relativistic field theory model of the deuteron~\cite{Ivanov95}.
This model has been constructed
in analogy with the $\sigma$-model and the extended Nambu-Jona-Lasinio
model~\cite{Itzykson80,Alfaro73,Nambu61}.

This result is due to our model approach using one-nucleon loop diagrams
for the 
description of strong low-energy interactions of the deuteron to other
particles. It is 
well-known that such fermion loop diagrams should possess
anomalies~\cite{Adler69,Bell69,Bardeen69,Gertsein69,Brown69}. As usually, 
these anomalies dominate the amplitudes of strong low-energy interactions of 
hadrons~\cite{Adler69,Wess71,Ivanov80,Ivanov81}. In the case of the
two-proton fusion we encounter the dominance of 
the anomaly of the AAV one-nucleon loop diagrams, that is the diagrams
having two 
axial-vector and one vector vertices. This is the reason why our result
cannot be
reduced to the obtained value in the potential approach.

In the low-energy limit when the 3-momenta of interacting protons tends to
zero the 
amplitude of the process 
p + p $\to$ D + ${\rm e}^+$ + $\nu_{\rm e}$ reads~\cite{Ivanov97}
\begin{eqnarray}\label{label1}
&&{\cal M}({\rm p} + {\rm p} \to {\rm D} + {\rm e}^+ \nu_{e}) = \nonumber\\
&&i\,C(v)\,g_{\rm A} M_{\rm N} \frac{G_{\rm F}}{\sqrt{2}}\,\frac{3g_{\rm V}}{4\pi^2}\,e^{\ast}_{\mu}(Q)\,[\bar{u}(k_{\nu})\gamma^{\mu}(1-\gamma^5)v(k_{\rm e})]\,\times\nonumber\\
&&\times\,\frac{\displaystyle \frac{g^2_{\pi\rm NN}}{8M^2_{\pi}}\Bigg(1 - \frac{8\sqrt{2}\pi\,M^2_{\pi}}{g^2_{\rm \pi NN}}\,\frac{a_{\rm S}}{M_{\rm N}}\Bigg)}{\displaystyle 1 - i\,\frac{g^2_{\pi\rm NN}}{8M^2_{\pi}}\Bigg(1 - \frac{8\sqrt{2}\pi\,M^2_{\pi}}{g^2_{\rm \pi NN}}\,\frac{a_{\rm S}}{M_{\rm N}}\Bigg)\,\frac{M^2_{\rm N}}{2\pi}\,v}\,[\bar{u^c}(p_1) \gamma^5 u(p_2)],
\end{eqnarray}
where $a_{\rm S}$ is the $^{1}{\rm S}_0$ pp scattering length the experimental value of which is $a_{\rm S} = ( -17.1\pm 0.2)\,{\rm fm}$~\cite{Ivanov97}. 

The cross section of the reaction 
p + p $\to$ D + ${\rm e}^+$ + $\nu_{\rm e}$ is given by 

\begin{eqnarray}\label{label2}
&&\sigma({\rm p} + {\rm p} \to {\rm D} + {\rm e}^+ + \nu_{\rm e}) =
\frac{C^2(v)}{v}\,\frac{9g^2_{\rm A}G^2_{\rm F}Q_{\rm D}}{1280 \pi^5}\,\,\varepsilon^5_{\rm D}\,M^3_{\rm N}f\Bigg(\frac{m_{\rm e}}{\varepsilon_{\rm D}}\Bigg) \times \nonumber\\
&&\times\,\frac{\displaystyle \Bigg[\frac{g^2_{\pi\rm NN}}{8M^2_{\pi}}\Bigg]^2\Bigg(1 - \frac{8\sqrt{2}\pi\,M^2_{\pi}}{g^2_{\rm \pi NN}}\,\frac{a_{\rm S}}{M_{\rm N}}\Bigg)^2}{\displaystyle 1 +\,\Bigg[\frac{g^2_{\pi\rm NN}}{8M^2_{\pi}}\Bigg]^2\Bigg(1 - \frac{8\sqrt{2}\pi\,M^2_{\pi}}{g^2_{\rm \pi NN}}\,\frac{a_{\rm S}}{M_{\rm N}}\Bigg)^2 \frac{M^4_{\rm N}}{4\pi^2}\,v^2}=\nonumber\\
&&= \frac{1}{\displaystyle 1 + 52743\,v^2}\,\times\,\frac{C^2(v)}{v}\,\times\,1.34\times10^{-48}\,{\rm cm}^2.
\end{eqnarray}

The cross section is calculated in units of $\hbar=c=1$. All
parameters in the above equations are defined in~\citeauthor{Ivanov97},
\citeyear{Ivanov97}.
The appearance of 
the factor $C(v)=\sqrt{2\pi\alpha/v}\exp(-\pi\alpha/v)$ taking into account
the Coulomb 
repulsion between protons at low energies  agrees with the result obtained by
\citeauthor{Bethe39}, \citeyear{Bethe39}. The reaction rate is then given
by~\cite{Ivanov97}

\begin{eqnarray}\label{label3}
&&<v\,\sigma({\rm p} + {\rm p} \to {\rm D} + {\rm e}^+ + \nu_{\rm e})> = 1.03\times 10^{-39}\,\times\,\frac{1}{\alpha}\,\times\,\frac{2}{\pi}\times\,\Bigg(\frac{1}{3}\Bigg)^2\times\nonumber\\
&&\times\,\sqrt{\frac{\pi}{3}}\,\times\,\frac{\displaystyle \tau^2\,e^{-\tau}}{\displaystyle 1 + 52743\,\frac{3\alpha^2 \pi^2}{\tau}} = 1.03\times 10^{-38}\,\frac{\displaystyle \tau^2\,e^{-\tau}}{\displaystyle 1 + \frac{83}{\tau}}\,\;{\rm cm}^3\,{\rm s}^{-1},
\end{eqnarray}

\noindent where $\tau$ is connected with the temperature~\cite{Bethe39}
\begin{eqnarray}\label{label4}
\tau = 3\Bigg(\frac{\alpha^2\pi^2 M_N}{4 k{\rm T}}\Bigg)^{1/3}.
\end{eqnarray}
The temperature dependence of Eq.~(3) coincides fully with that derived by 
\citeauthor{Bethe39}, \citeyear{Bethe39}.
Setting $ T = T_{\rm c}=15.5 \times 10^6$\,K, where
$T_{\rm c}$ is the temperature of the solar core in 
the Standard Solar model~\cite{Rolfs88}, we get $\tau=13.56$, and obtain
the following estimate
\begin{eqnarray}\label{label5}
<v\,\sigma({\rm p} + {\rm p} \to {\rm D} + {\rm e}^+ + \nu_{\rm e})> =
 3.44\times 10^{-43}\,{\rm cm}^3\,{\rm s}^{-1}\,.
\end{eqnarray}
This value is by a factor of 2.9 larger than the one calculated within the
potential 
approach~(\citeauthor{Rolfs88}, \citeyear{Rolfs88}, see also
\citeauthor{Bahcall89}, \citeyear{Bahcall89}
and \citeauthor{Kamionkowski94}, \citeyear{Kamionkowski94}). 
The magnitude of the theoretical uncertainty of the relativistic field
theory model of 
the deuteron is expected of order $\Delta =\pm 30\%$~\cite{Ivanov97}.

The enhancement of the amplitude of the p + p $\to$ D + W transition found
in our 
approach is related to the computation of the amplitude in terms of one-nucleon 
loop diagrams. Indeed, the structure function in the momentum
representation~\cite{Ivanov97}
defining the effective Lagrangian of the p + p $\to$ D + W transition is due 
to the contribution of the anomalous part of the AAV one-nucleon loop
diagram~\cite{Wess71,Ivanov80,Ivanov81}. 
Such an anomalous contribution, produced by vacuum fluctuations of virtual
nucleons, 
has a quantum-field-theory nature and cannot be obtained within the potential 
approach describing strong low-energy interactions of the protons and the
deuteron 
in terms of the overlap integral of the wave functions of two protons and the 
deuteron. The ambiguity of the computation of the AAV-anomaly, produced by
the shift 
of the virtual nucleon momentum, has been fixed by the requirement of gauge
invariance 
under gauge transformations of the deuteron field~\cite{Ivanov97}. This is
very similar to the 
removal of the ambiguity appearing for the computation of the
Adler-Bell-Jackiw-Bardeen
anomaly~\cite{Adler69,Bell69,Bardeen69}.

\begin{table}
\caption{Contributions from the main components of the neutrino 
fluxes (in SNU) for the gallium and chlorine
$S_{\rm Ga}$ and $S_{\rm Cl}$ detectors, respectively,
according to a SSM and our alternative solar model (ASM).
In the last line the summed up neutrino fluxes are compared with the
corresponding experimental data.
In the last column the parameters $\alpha_i$ of the power-law
behavior are shown. The errors are due to the assumed 30\,\% uncertainty
of the reaction rate for p + p $\to$ D + $e^{\,+}$ + $\nu_{\rm e}$.}
\begin{center}
\begin{tabular}{cccccccc}\hline
& \multicolumn{3}{c}{$S_{\rm Ga}$}&
\multicolumn{3}{c}{$S_{\rm Cl}$}& \\\cline{2-7}
& SSM$^{\rm a}$ & our model& experiment$^{\rm b}$ &
SSM$^{\rm b}$ & ASM & experiment$^{\rm c}$ &
\raisebox{1.5ex}[-1.5ex]{$\alpha_i^{\rm d}$} \\\hline
pp & 69.7 & 75.1$^{+1.1}_{-1.9}$ &  & 0.00 & 0.00 & &0.07 \\
pep & 3.0 & 3.2$^{+0,0}_{-0.0}$ &  & 0.22 & 0.24$^{+0.01}_{-0.01}$ & & 0.07 \\
$^7{\rm Be}$ & 37.7 & 11.7$^{-2,9}_{+5.6}$ &  & 1.24 & 0.39$^{-0.10}_{+0.17}$ & &$-1.1$ \\
$^{13}{\rm N}$ & 3.8 & 0.4$^{-0.2}_{+0.4}$ &  & 0.11 & 0.01$^{-0.01}_{+0.01}$ & &$-2.2$ \\
$^{15}{\rm O}$ & 6.3 & 0.6$^{-0.3}_{+0.6}$ &  & 0.37 & 0.04$^{-0.01}_{+0.04}$ & &$-2.2$ \\
$^{8}{\rm B}$ & 16.1 & 0.9$^{-0.4}_{+1.5}$ &  & 7.36 &
0.42$^{-0.22}_{+0.67}$ & &$-2.7$ \\ \hline
Sum & 136.6 & 91.9$^{-2.9}_{+6.2}$ & $77.1 \pm 13.4$ & 9.30 & 1.10$^{-0.33}_{+0.88}$
& $2.55\pm 0.35$ &\\ \hline
\end{tabular}
\end{center}

{\footnotesize
$^{\rm a}$ \citeauthor{Bahcall95}, \citeyear{Bahcall95}\\
$^{\rm b}$ \citeauthor{GALLEX96}, \citeyear{GALLEX96}\\
$^{\rm c}$ \citeauthor{Cleveland95}, \citeyear{Cleveland95}\\
$^{\rm d}$ \citeauthor{Castellani97}, \citeyear{Castellani97}
}
\end{table}

\begin{table}
\caption{Contributions from the main components of the neutrino 
fluxes (in 10$^6$\,cm$^{-2}$\,s$^{-1}$) for the KAMIOKANDE detector
according to a SSM and our alternative solar model (ASM).
In the last line the neutrino flux is compared with the
corresponding experimental data.
In the last column the parameter $\alpha_i$ of the power-law
behavior is shown. The errors are due to the assumed 30\,\% uncertainty
of the reaction rate for p + p $\to$ D + $e^{\,+}$ + $\nu_{\rm e}$.}
\begin{center}
\begin{tabular}{ccccc}\hline
& \multicolumn{3}{c}{$S_{\rm Kam}$}& \\\cline{2-4}
& SSM$^{\rm a}$ & ASM & experiment$^{\rm b}$ &
\raisebox{1.5ex}[-1.5ex]{$\alpha_i^{\rm c}$} \\\hline
$^{8}{\rm B}$ & 6.62 & $0.37^{-0.19}_{+0.61}$ & & -2.7 \\
Sum & 6.62 & $0.37^{-0.19}_{+0.61}$ & 2.73$\pm$0.51 &\\\hline
\end{tabular}
\end{center}

{\footnotesize
$^{\rm a}$ \citeauthor{Bahcall95}, \citeyear{Bahcall95}\\
$^{\rm b}$ \citeauthor{KAMIOKANDE95}, \citeyear{KAMIOKANDE95}\\
$^{\rm c}$ \citeauthor{Castellani97}, \citeyear{Castellani97}
}
\end{table}

\section{Discussion}

One of the possible relaxations of the solar neutrino problem is to lower
the temperature in the
center of the sun in comparison to that predicted by SSM's as
$T_{\rm c} = 15.5\times 10^6\,{\rm K}$~\cite{Rolfs88}. Indeed, due to
strong dependence of the  solar neutrino fluxes on $T_{\rm c}$  just a 20
to 30\,\%
diminishing of $T_{\rm c}$ 
leads to a suppression of the neutrino fluxes by more than an order of
magnitude~\cite{Castellani97}.
In order to reduce $T_{\rm c}$  one can resort to the change of physical
and chemical 
phenomenological inputs which determine the structure of the
star~\cite{Castellani97}.
A process  which influences substantially the temperature in the center of
the sun is the 
reaction p + p $\to$ D + e$^+$ + $\nu_{\rm e}$. The 
magnitude of the reaction rate for p + p $\to$ D + e$^+$ + $\nu_{\rm e}$ is
directly
related to the solar luminosity that must be reproduced by any solar
model. Therefore, an enhancement of 
the cross section magnitude of two-proton fusion leads to a decrease of 
$ T_c$ that suppresses the solar neutrino fluxes for
the high-energy solar neutrinos~\cite{Castellani97}.
In order to reconcile the enhancement of our new reaction rate for
p + p $\to$ D + $e^{\,+}$ + $\nu_{\rm e}$ by the factor of 2.9 with the solar
luminosity we must assume that the temperature in the solar core equals 
$T_{\rm c}= 13.8^{-0.4}_{+0.5} \times 10^6\,{\rm K}$~\cite{Castellani97}.

The solar neutrino fluxes $\Phi_i$, where i = pp, pep, $^7$Be, $^{13}$N,
$^{15}$O
and $^8$B, can be represented in the form of a power-law behavior
\cite{Castellani97}, i.e.,
\begin{eqnarray}\label{label6}
\Phi_i\,=\,x^{\alpha_i}\,\Phi^*_i.
\end{eqnarray}
$\Phi^*_i$ is the neutrino flux, predicted by SSM's, and the parameter $x$
in our
definition reads
\begin{eqnarray}\label{label7}
x\,=\,\frac{<v\,\sigma({\rm p} + {\rm p} \to {\rm D} + {\rm e}^+ +
\nu_{\rm e})>}{<v\,\sigma({\rm p} + {\rm p} \to {\rm D} +
{\rm e}^+ + \nu_{\rm e})>^*}=2.90 \pm 0.87,
\end{eqnarray}
where $<v\,\sigma({\rm p} + {\rm p} \to {\rm D} + {\rm e}^+ +
\nu_{\rm e})>^*$ is the quantity calculated in the potential approach.

The values of the parameters ${\alpha}_{\rm i}$ are given in Tables II and III
and can be found in Table X of \citeauthor{Castellani97},
\citeyear{Castellani97}.
In these tables we also show the
neutrino fluxes that should give the contributions to the signals detected
in the gallium and chlorine and KAMIOKANDE neutrino
detectors. We have normalized our predictions to
the results obtained with a reference SSM~\cite{Bahcall95}.

It is seen that our alternative solar model (ASM)
explains reasonably well the experimental data of
the gallium experiments (see Table II, last line). For the neutrino flux
measured
in the chlorine
experiment our model prediction is too small by about a factor of 2
(see Table II, last line).
One can see that our prediction for the solar neutrino flux
is much too small when compared with the KAMIOKANDE experimental data (see
Table III,
last line).

Another possibility to compare the results of the neutrino detectors with
solar models is in terms of the neutrino fluxes for the Be-
and B neutrinos.
Roughly speaking the experimental results of the three neutrino
detectors imply for the different neutrino fluxes when compared
to a SSM (see
Tables II and III):

\begin{eqnarray}\label{8}
\Phi_{\rm pp} & \approx & \Phi_{\rm pp}^{\rm SSM}\\
\Phi_{\rm Be+CNO} & \ll & \Phi_{\rm Be+CNO}^{\rm SSM}\\
\Phi_{\rm B} & \approx & 0.35\,\Phi_{\rm B}^{\rm SSM}
\end{eqnarray}
\noindent
Therefore, the solar neutrino problem can also be formulated
when compared to SSM's as the problem of the missing
Be-neutrinos and the reduced flux of the B-neutrinos.

Using our new reaction rate for p + p $\to$ D + e$^+$ + $\nu_{\rm e}$
we obtain the following approximate relationship between
the experimental results of the three neutrino
detectors and our ASM:

\begin{eqnarray}\label{9}
\Phi_{\rm pp} & \approx & \Phi_{\rm pp}^{\rm ASM}\\
\Phi_{\rm Be+CNO} & \approx & \Phi_{\rm Be+CNO}^{\rm ASM}\\
\Phi_{\rm B} & \approx & 7\,\Phi_{\rm B}^{\rm ASM}
\end{eqnarray}
\noindent
Therefore, the solar neutrino problem can also be formulated
when compared to the ASM as the problem of the excessive B-neutrinos.

A more quantitative calculation gives the following constraints between
the detected solar neutrino signals and the B- and (Be+CNO)-fluxes
(see Eq.~(61) of \citeauthor{Castellani97}, \citeyear{Castellani97}):

\begin{eqnarray}\label{10}
S_{\rm Ga} & = & 80.1 + 6.14 \Phi_{\rm Be+CNO} + 2.43 \Phi_{\rm B}\\
S_{\rm Cl} & = & 0.248 + 0.236 \Phi_{\rm Be+CNO} + 1.11 \Phi_{\rm B}\\
S_{\rm Ka} & = & \Phi_{\rm B}
\end{eqnarray}

Here $S_{\rm Ga}$ and $S_{\rm Cl}$ are given in SNU,
$\Phi_{\rm Be+CNO}$ in 10$^9$\,cm$^{-2}$\,s$^{-1}$, and
$S_{\rm Ka}$ and $\Phi_{\rm B}$ in 10$^6$\,cm$^{-2}$\,s$^{-1}$.
The flux of the pp-neutrinos has been eliminated in the above equations by using
the solar luminosity constraint, i.e., by assuming that the flux
of the pp-neutrinos is constrained approximately by the solar luminosity.
The result for each experiment can then be plotted in the
($\Phi_{\rm B},\Phi_{\rm Be+CNO}$)-plane as shown in Fig.~3.
Also in this
figure the values of these neutrino fluxes obtained
in a standard solar model (SSM)~\cite{Bahcall95} as well as in our
alternative solar model (ASM) with the new reaction rate are shown.

\begin{table}
\caption{Differences of the neutrino fluxes predicted by
SSM's and our ASM from the values observed by the different
neutrino detectors in standard experimental deviations $\sigma$
of the corresponding experiments.}
\begin{center}
\begin{tabular}{ccc}\hline
Neutrino detector & Deviation of SSM in $\sigma$ & Deviation of ASM in
$\sigma$\\\hline
Gallium detector  & +8 & +3 \\
Chlorine detector & +19 & $-4$ \\
KAMIOKANDE & +8 & $-5$ \\\hline
\end{tabular}
\end{center}
\end{table}

In Table IV the differences of the neutrino fluxes predicted by the
SSM and the ASM from the values observed by different
neutrino detectors derived from Tables II and III are shown
using the standard deviations $\sigma$
of the corresponding experiments. These differences are considerable less
in the ASM than in the SSM. A compelling argument for a resolution
in terms of new particle physics should rest on a dramatic discrepancy
(often estimated at 5\,$\sigma$) between the neutrino detectors
and the flux predictions of solar models~\cite{Cumming96}. As
can be seen from Table IV such a discrepancy is barely reached when
assuming our ASM.

Helioseismological observations are an important tool to investigate
the solar interior dynamics and structure with great
precision~\cite{Christensen-Daalgaard96}.
Interesting enough the sound speed derived from SSM's including
element diffusion agrees with helioseismological measurements
with a precision better than 0.2\,\%~\cite{Bahcall97}. Even tiny fractional
changes in the temperature and the molecular weight would
produce measurable discrepancies in the precisely determined
helioseismological sound speed. Since the temperature in the
solar center for our ASM is about 9\,\% lower than in the SSM, it remains
questionable if the sound speed can also be reproduced by the ASM.
Investigations in this direction
are in progress.

\acknowledgements
This work was partially supported  by the Fonds zur F\"orderung der
wissenschaftlichen Forschung in \"Osterreich (project P10361--PHY).

\newpage

\section*{Figure captions}

Fig.~1: The three solar pp-chains

\noindent
Fig.~2: Spectrum of the solar neutrinos. At the top of the figure
the thresh\-olds of the different neutrino detectors are indicated:\\
Ga \ldots gallium detectors, Cl \ldots chlorine detector,
K II \ldots KAMIOKANDE detector.

\noindent
Fig.~3: The $^8$B and $^7$Be neutrino fluxes, consistent with the
luminosity constraint and experimental results. The shaded areas correspond
to the (best-fit $\pm$ 1\,$\sigma$) experimental values for the chlorine,
gallium and KAMIOKANDE neutrino detectors.
The full circles indicate the predictions by a standard solar model (SSM)
and our alternative solar model (ASM).

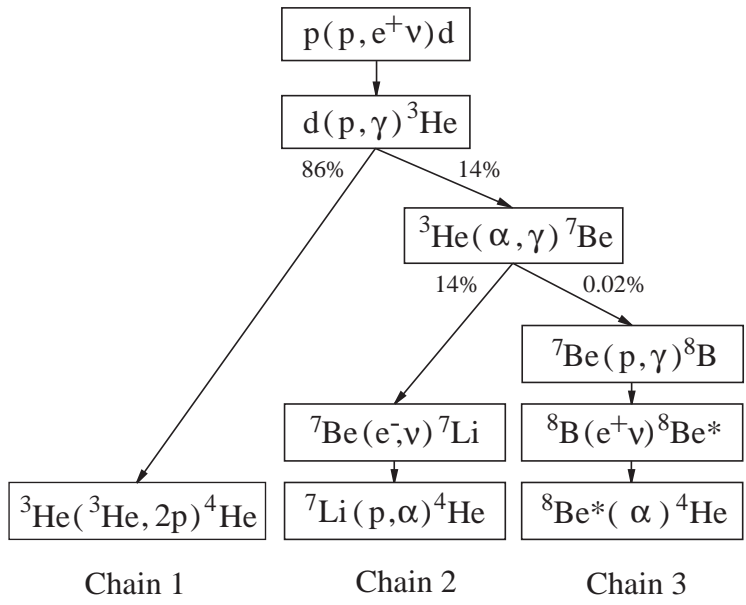
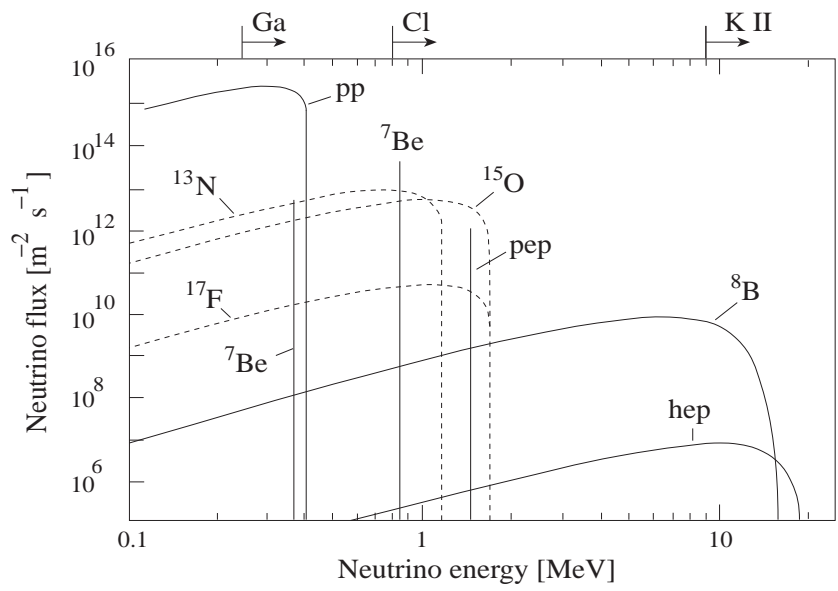
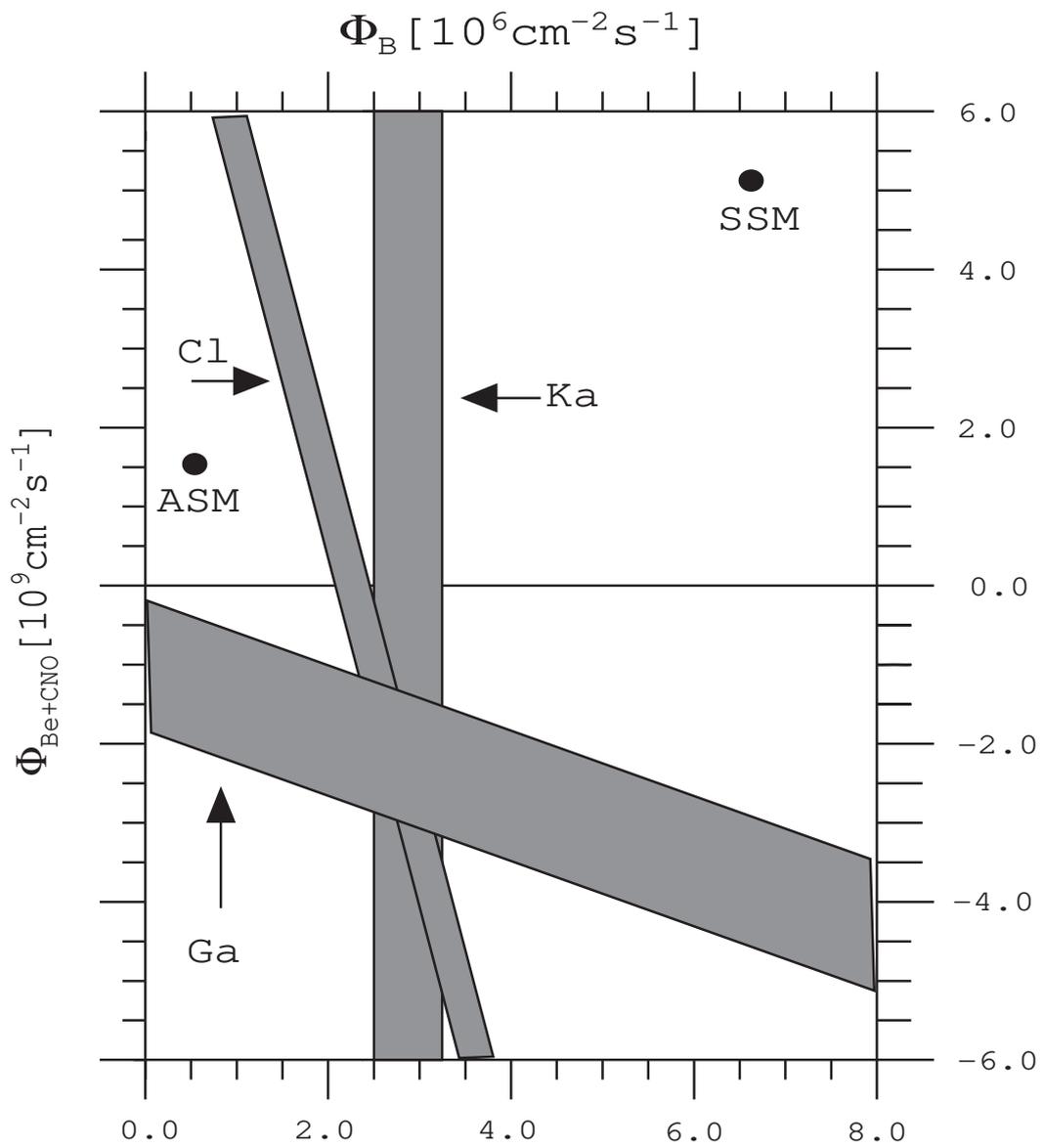
\end{document}